# Dust-acoustic waves and stability in the permeating dusty plasma: I. Maxwellian distribution

Jingyu Gong[1], Zhipeng Liu[1,2], and Jiulin Du[1, a)]

[1]*Department of Physics, School of Science, Tianjin University, Tianjin 300072, China*

[2]*Department of Fundamental Subject, Tianjin Institute of Urban Construction, Tianjin 300384, China*

**Abstract**

The dust-acoustic waves and their stability in the permeating dusty plasma with the Maxwellian velocity distribution are investigated. We derive the dust-acoustic wave frequency and instability growth rate in two limiting physical cases that the thermal velocity of the flowing dusty plasma is (a) much larger than, and (b) much smaller than the phase velocity of the waves. We find that the stability of the waves depend strongly on the velocity of the flowing dusty plasma in the permeating dusty plasma. The numerical analyses are made based on the example that a cometary plasma tail is passing through the interplanetary space plasma. We show that, in case (a), the waves are generally unstable for any flowing velocity, but in case (b), the waves become unstable only when the wave number is small and the flowing velocity is large. When the physical conditions are between these two limiting cases, we gain a strong insight into the dependence of the stability criterions on the physical conditions in the permeating dusty plasma.

---

a) Corresponding author, E-mail address: jiulindu@yahoo.com.cn

## I. INTRODUCTION

Dusty plasma contains electrons, ions and additional charged components of micron- or submicron-sized particulates often known as dust grains.[1-5] The dust grains may acquire charges through a variety of interactions with electrons and ions.[6,7] These interactions can also take place between the dust grains and they can make a response to an external perturbation. The dusty plasmas exist ubiquitously in the space regions such as the interstellar clouds, the circumstellar clouds, the interplanetary space, the comets, the planetary rings, the Earth's atmosphere, and the lower ionosphere, etc. These space plasmas are generally in motion, and thus an interpenetrating (inter-permeating) plasma physical phenomenon may frequently occur if two or more space plasmas encounter in the same space region. This is the so-called permeating plasma. A model of the permeating plasma can be considered to consist of two parts, called the flowing plasma and the target plasma. The target plasma is in a relative static state, while the flowing plasma has a flowing velocity. And the flowing plasma can moves through the target plasma when the two plasmas encounter in the space. An example of the permeating plasma is that the solar/stellar wind plasma is inter-permeating with the surrounding cometary plasma.[8,9] Usually, the solar/stellar wind plasma is an electron-ion plasma, while the cometary plasma is a dusty plasma. Occasionally, when a long-period comet's dust trail showers the earth, the dusty plasmas in the earth's atmosphere will encounter the cometary dust trails to form the permeating dusty plasma.[10,11] Since the interplanetary space is full of dust grains, the physical situation of permeating dusty plasma can often appear in the space where the cometary plasma tails encounter with the interplanetary medium plasmas[1]. Therefore, the permeating dusty plasma is a basic physical phenomenon in space plasma, in which the waves and their stability are important for us to explore the understanding of a variety of space physical phenomena.

The diverse investigations of ion-acoustic waves,[12-17] electron-acoustic waves[18,19] and dust-acoustic waves[20-26] driven by the space plasmas have attracted long-term and growing interests. As early as 1990, dust-acoustic waves in dusty plasma were investigated by a theoretical method,[27] and later, the dust-acoustic waves were studied by experimental methods.[28,29] Most recently, a current-less kinetic instability for ion- and dust-acoustic waves was analyzed in the permeating dusty plasmas.[8,9] In all these works, the employed statistical distributions include the Maxwellian distribution in the traditional statistics and the power-law $q$-distribution[30] in nonextensive statistics. In this work, we will generally investigate the dust-acoustic waves and their stability in the permeating dusty plasma with the Maxwellian velocity distributions. And in two limiting physical cases that the thermal velocity of the flowing dusty plasma is much larger than, and much smaller than the phase velocity of the waves, we give our theoretical results and numerical calculations. In paper II[31], we will investigate these characteristics in the permeating dusty plasma with the power-law distributions, where the nonextensive effects[32] in the nonequilibrium plasmas are considered.

The paper is organized as follows. In Sec. II, the basic theory for dust-acoustic waves in the plasma with the Maxwellian velocity distribution is reviewed. In Sec. III, the frequency and instability growth rate of the dust-acoustic waves are investigated in the permeating plasma. In Sec. IV, Numerical analyses of the dust-acoustic wave instability driven by a cometary plasma tail passing through the interplanetary space are presented. Finally, the conclusions are given in Sec. V.

## II. BASIC THEORY OF THE DUST-ACOUSTIC WAVES IN THE DUSTY PLASMA WITH MAXWELLIAN DISTRIBUTION

In this section, we make a brief review on the basic theory and the dispersion function for the dust-acoustic waves in a collisionless, unbounded and unmagnetized dusty plasma. The kinetic equation for the plasma is the linearized Vlasov equation,[33]

$$\frac{\partial f_{\alpha 1}(r,v,t)}{\partial t} + v \cdot \nabla f_{\alpha 1}(r,v,t) + \frac{Q_\alpha}{m_\alpha} E_1 \cdot \nabla_v f_{\alpha 0}(r,v) = 0, \quad (1)$$

where the subscript $\alpha = d, i, e$ denotes the dust grains, the ions and the electrons, respectively; $f_{\alpha 1}(r,v,t)$ is a distribution function of the plasma at time $t$, position $r$ and velocity $v$, which here represents only a small perturbation about the equilibrium distribution function $f_{\alpha 0}(\boldsymbol{r},\boldsymbol{v})$; $Q_\alpha$ is a charge of the plasma component $\alpha$, and $\boldsymbol{E}_1$ is a small perturbation electric field, satisfying the linearized Poisson equation:

$$\nabla \cdot \boldsymbol{E}_1 = \frac{1}{\varepsilon_0}\sum_\alpha Q_\alpha \int f_{\alpha 1} d\boldsymbol{v}. \qquad (2)$$

Considering $f_{\alpha 1}(\boldsymbol{r},\boldsymbol{v},t) \propto \exp[i(\boldsymbol{k}\cdot\boldsymbol{r}-\omega t)]$ and $\boldsymbol{E}_1 \propto \exp[i(\boldsymbol{k}\cdot\boldsymbol{r}-\omega t)]$, and then making the Fourier transformation for $\boldsymbol{r}$ and the Laplace transformation for $t$ in Eqs.(1) and (2) respectively, one can obtain

$$-i\omega f_{\alpha 1} + i\boldsymbol{k}\cdot\boldsymbol{v} f_{\alpha 1} + \frac{Q_\alpha}{m_\alpha}\boldsymbol{E}_1 \cdot \nabla_v f_{\alpha 0} = 0, \qquad (3)$$

$$i\boldsymbol{k}\cdot\boldsymbol{E}_1 = \frac{1}{\varepsilon_0}\sum_\alpha Q_\alpha \int f_{\alpha 1} d\boldsymbol{v}. \qquad (4)$$

Let the wave vector $\boldsymbol{k}$ be along x-axis and the velocity is $v_x = u$. According to the Landau path integral,[33] the dispersion relation, also called polarizability, reads

$$\varepsilon(\omega,k) = 1 + \sum_\alpha \chi_\alpha = 0, \qquad (5)$$

where the physical quantity $\chi_\alpha$ is

$$\chi_\alpha = \frac{\omega_{p\alpha}^2}{k^2} \int \frac{\partial \hat{f}_{\alpha 0}/\partial u}{\omega/k - u} du, \qquad (6)$$

$\omega_{p\alpha} = \sqrt{n_{\alpha 0} Q_\alpha^2/\varepsilon_0 m_\alpha}$ is the naturally oscillating frequency of plasma, and $\hat{f}_{\alpha 0} = f_{\alpha 0}/n_{\alpha 0}$ is the normalized equilibrium distribution function. Generally speaking, if the permeating dusty plasma is near thermal equilibrium, the particles for the plasma component $\alpha$ obey approximately the Maxwellian velocity distribution,

$$\hat{f}_{\alpha 0} = \frac{1}{\sqrt{2\pi} v_{T\alpha}} \cdot \exp\left[-\frac{(v-v_{\alpha 0})^2}{v_{T\alpha}^2}\right], \qquad (7)$$

with the thermal velocity $v_{T\alpha} = \sqrt{2k_B T_\alpha/m_\alpha}$. Thus, the dispersion function can be expressed[33] by

$$Z(\xi_\alpha) = \frac{1}{\sqrt{\pi}} \int_{-\infty}^{\infty} \frac{\exp(-x^2)}{x - \xi_\alpha} dx, \qquad (8)$$

where $\xi_\alpha = (v_\phi - v_{\alpha 0})/v_{T\alpha}$, and $v_\phi = \omega/k$ is the phase velocity of the dust-acoustic waves. The integrand in Eq.(8) has a singular point at $x = \xi_\alpha$. According to the Landau path integral,[33] Eq.(8) can be written as

$$Z(\xi_\alpha) = \frac{1}{\sqrt{\pi}} \Pr \int_{-\infty}^{\infty} \frac{\exp(-x^2)}{x - \xi_\alpha} dx + i\sqrt{\pi} \exp(-\xi_\alpha^2), \qquad (9)$$

where the real part is

$$\mathrm{Re}[Z(\xi_\alpha)] = \begin{cases} -\dfrac{1}{\xi_\alpha} - \dfrac{1}{2\xi_\alpha^3} - \dfrac{3}{4\xi_\alpha^5} - \cdots, & (\xi_\alpha \gg 1) \\ -2\xi_\alpha \left(1 - \dfrac{2}{3}\xi_\alpha^2 + \cdots\right), & (\xi_\alpha \ll 1) \end{cases} \qquad (10)$$

Then, the dispersion relation, Eq.(5), can be expressed with the dispersion function as

$$\varepsilon(\omega, k) = 1 + \sum_\alpha \frac{1}{k^2 \lambda_{D\alpha}^2} [1 + \xi_\alpha Z(\xi_\alpha)] = 0, \qquad (11)$$

where $\lambda_{D\alpha} = v_{T\alpha}/\sqrt{2}\omega_{p\alpha}$ is the Debye length of the component $\alpha$ in the dusty plasma.

## III. DUST-ACOUSTIC WAVE FREQUENCY AND INSTABILITY GROWTH RATE IN THE PERMEATING DUSTY PLASMA

In the permeating dusty plasma, one part is the flowing dusty plasma and the other part is the relatively static target dusty plasma. Each part should both satisfy the quasi-neutral conditions, i.e. $Q_{jd0} n_{jd0} = Q_{ji0} n_{ji0} - Q_{je0} n_{je0}$ for $j$th part, where $j = f, s$ denote the flowing dusty plasma ($f$) and the target dusty plasma ($s$); $Q_{jd0}$, $Q_{ji0}$ and $Q_{je0}$ are the unperturbed charges, and $n_{jd0}$, $n_{ji0}$ and $n_{je0}$ are the unperturbed number densities of the dust grains, the ions and the electrons, respectively.

In order to discuss the dust-acoustic waves and their stability in the permeating dusty plasma, we appoint a coordinate system which is along with the target dusty plasma. And so we can assume that the target dusty plasma is static (or moves with a constant velocity $\mathbf{v}_{s0} = v_{s0}\mathbf{e}_z$ in the $z$ direction) in this coordinate system, while the flowing dusty plasma is in motion with a velocity $\mathbf{v}_{f0} = v_{f0}\mathbf{e}_z$. Without loss of generality, we let $v_{s\alpha} = 0$ and $v_{f\alpha} = v_{f0}$ for each plasma component $\alpha$. In general, the following relations between the thermal velocity (with the subscript $T$) and the phase velocity (with the subscript $\phi$) of the waves may be satisfied, i.e. $v_{Tsd} \ll v_\phi \ll v_{Tsi}, v_{Tse}$ and $v_\phi - v_{f0} \ll v_{Tfi}, v_{Tfe}$, which will simplify our theoretical calculations. However, the relation between $v_\phi - v_{f0}$ and $v_{Tfd}$ is uncertain. To get a more clear result, we need to give the discussions in two limiting physical cases (a) and (b) for this relation.

(*a*). In the case of $v_\phi - v_{f0} \ll v_{Tfd}$, i.e. the thermal velocity of flowing dusty plasma is much larger than the phase velocity of the waves, we have $\xi_{sd} \gg 1$ and $\xi_{si}, \xi_{se}, \xi_{fd}, \xi_{fi}, \xi_{fe} \ll 1$, and then we can expand Eq.(11) as

$$\varepsilon(\omega, k) \approx 1 + \frac{1}{k^2 \lambda_D^2} - \frac{\omega_{psd}^2}{\omega^2} - \frac{3k^2 v_{Tsd}^2 \omega_{psd}^2}{2\omega^4}$$
$$+ i2\sqrt{\pi} \left[ \frac{\omega \omega_{psd}^2}{k^3 v_{Tsd}^3} \exp\left(-\frac{\omega^2}{k^2 v_{Tsd}^2}\right) + \frac{\omega \omega_{psi}^2}{k^3 v_{Tsi}^3} + \frac{\omega \omega_{pse}^2}{k^3 v_{Tse}^3} \right. \quad (12)$$
$$\left. + (\omega - kv_{f0}) \left( \frac{\omega_{pfd}^2}{k^3 v_{Tfd}^3} + \frac{\omega_{pfi}^2}{k^3 v_{Tfi}^3} + \frac{\omega_{pfe}^2}{k^3 v_{Tfe}^3} \right) \right]$$

where $1/\lambda_D^2 = 1/\lambda_{Dsi}^2 + 1/\lambda_{Dse}^2 + 1/\lambda_{Dfd}^2 + 1/\lambda_{Dfi}^2 + 1/\lambda_{Dfe}^2$. Following the Landau's path integral, if the wave vector $\mathbf{k}$ is a real number and the frequency is $\omega = \omega_r + i\gamma$, the real part $\omega_r = \mathrm{Re}\,\omega$ is the frequency of the waves, while the imaginary part $\gamma = \mathrm{Im}\,\omega$ is the growth rate of the waves in the Landau damping. If there is $\gamma > 0$, the waves will grow and they becomes unstable.

In the case of weak damping, the growth rate is small, $\gamma \ll \omega_r$. The dielectric constant $\varepsilon(\omega_r + i\gamma, k)$ can be expressed as a series for the growth rate,[33]

$$\varepsilon(\omega,k) \approx \varepsilon_r(\omega_r,k) + i\gamma \left.\frac{\partial \varepsilon_r}{\partial \omega_r}\right|_{\omega=\omega_r} + i\varepsilon_i(\omega_r,k). \tag{13}$$

Let the real part of Eq.(13) be zero, i.e., $\varepsilon_r(\omega,k) \approx \varepsilon_r(\omega_r,k) = 0$. Combining with Eq.(12), one can get the frequency of dust-acoustic waves,

$$\omega_r \approx \frac{\omega_{psd} k \lambda_D}{\sqrt{1+k^2 \lambda_D^2}}, \tag{14}$$

where the small quantities have been neglected. If one let the imaginary part of Eq.(13) be zero, one can get the instability growth rate of the waves,

$$\gamma = -B\left[C\left(1 - \frac{v_{f0}}{v_\phi}\right) + D\right], \tag{15}$$

with the parameters

$$B = \frac{\sqrt{\pi}\omega_r^3}{\omega_r^2 + 3k^2 v_{Tsd}^2} \cdot \frac{\omega_r^3}{k^3 v_{Tsd}^3}, \tag{16}$$

$$C = \frac{v_{Tsd}^3 \omega_{pfd}^2}{v_{Tfd}^3 \omega_{psd}^2} + \frac{v_{Tsd}^3 \omega_{pfi}^2}{v_{Tfi}^3 \omega_{psd}^2} + \frac{v_{Tsd}^3 \omega_{pfe}^2}{v_{Tfe}^3 \omega_{psd}^2}, \tag{17}$$

and

$$D = \exp\left(-\frac{\omega_r^2}{k^2 v_{Tsd}^2}\right) + \frac{v_{Tsd}^3 \omega_{psi}^2}{v_{Tsi}^3 \omega_{psd}^2} + \frac{v_{Tsd}^3 \omega_{pse}^2}{v_{Tse}^3 \omega_{psd}^2}. \tag{18}$$

If the growth rate is $\gamma > 0$, the dust-acoustic waves become unstable. From Eq.(15), one therefore finds the stability condition,

$$v_{f0} < v_\phi \left(1 + \frac{D}{C}\right). \tag{19}$$

Obviously, for the dust-acoustic wave instability, there is a critical velocity of the flowing dusty plasma: $u_{f0} \equiv v_\varphi(1 + D/C)$. Namely, if $v_{f0} > u_{f0}$, the dust-acoustic waves are unstable. In reverse, if $v_{f0} < u_{f0}$, the growth rate is $\gamma < 0$ and then the dust-acoustic waves are stable. But if $v_{f0} = u_{f0}$, the waves are at the critical stability.

(b). In the case of $v_{Tfd} \ll v_\phi - v_{f0}$, i.e. the thermal velocity of flowing dusty plasma is much smaller than the phase velocity of the waves, we have $\xi_{sd} \gg 1$, $\xi_{fd} \gg 1$ and $\xi_{si}, \xi_{se}, \xi_{fi}, \xi_{fe} \ll 1$. In this case, Eq.(11) is expanded as

$$\varepsilon(\omega,k) \approx 1 + \frac{1}{k^2\lambda_D^2} - \frac{\omega_{psd}^2}{\omega^2} - \frac{3k^2 v_{Tsd}^2 \omega_{psd}^2}{2\omega^4} - \frac{\omega_{pfd}^2}{(\omega-kv_{f0})^2} - \frac{3k^2 v_{Tsd}^2 \omega_{psd}^2}{2(\omega-kv_{f0})^4}$$

$$+ i2\sqrt{\pi} \left\{ \frac{\omega \omega_{psd}^2}{k^3 v_{Tsd}^3} \exp\left(-\frac{\omega^2}{k^2 v_{Tsd}^2}\right) + \frac{\omega \omega_{psi}^2}{k^3 v_{Tsi}^3} + \frac{\omega \omega_{pse}^2}{k^3 v_{Tse}^3} + \right.$$

$$\left. + (\omega - kv_{f0}) \left( \frac{\omega_{pfd}^2}{k^3 v_{Tfd}^3} \exp\left[-\frac{(\omega-kv_{f0})^2}{k^2 v_{Tfd}^2}\right] + \frac{\omega_{pfi}^2}{k^3 v_{Tfi}^3} + \frac{\omega_{pfe}^2}{k^3 v_{Tfe}^3} \right) \right\} \quad (20)$$

where $1/\lambda_D^2 = 1/\lambda_{Dsi}^2 + 1/\lambda_{Dse}^2 + 1/\lambda_{Dfi}^2 + 1/\lambda_{Dfe}^2$. The frequency $\omega_r$ of the dust-acoustic waves can be determined by setting the real part of Eq.(20) to be zero,

$$1 + \frac{1}{k^2\lambda_D^2} - \frac{\omega_{psd}^2}{\omega_r^2} - \frac{\omega_{pfd}^2}{(\omega_r - kv_{f0})^2} = 0, \quad (21)$$

where we have neglected the two terms containing $v_{Tsd}^2$ in the case of $v_{Tsd} \ll v_\phi$ and $v_{Tfd} \ll (v_\phi - v_{f0})$. For the low frequency acoustic mode, $\omega_r \ll kv_{f0}$, we obtain

$$\omega_r \approx \frac{\omega_{psd} k \lambda_D'}{\sqrt{1 + k^2 \lambda_D'^2}}. \quad (22)$$

By setting the imaginary part of Eq.(20) to be zero, the instability growth rate of the wave is determined as

$$\gamma = -B' \left[ C' \left(1 - \frac{v_{f0}}{v_\phi}\right) + D' \right], \quad (23)$$

with the parameters

$$B' = \frac{\sqrt{\pi} \omega_r^4 / k^3 v_{Tsd}^3}{1 + \frac{3k^2 v_{Tsd}^2}{\omega_r^2} + \frac{\omega_{pfd}^2}{\omega_{psd}^2} \frac{\omega_r^3}{(\omega_r - kv_{f0})^3} \left[1 + \frac{3k^2 v_{Tfd}^2}{(\omega_r - kv_{f0})^2}\right]}, \quad (24)$$

$$C' = \frac{v_{Tsd}^3 \omega_{pfd}^2}{v_{Tsi}^3 \omega_{psd}^2} \exp\left[-\frac{(\omega_r - kv_{f0})^2}{k^2 v_{Tfd}^2}\right] + \frac{v_{Tsd}^3 \omega_{pfi}^2}{v_{Tfi}^3 \omega_{psd}^2} + \frac{v_{Tsd}^3 \omega_{pfe}^2}{v_{Tfe}^3 \omega_{psd}^2}, \quad (25)$$

and

$$D' = \exp\left(-\frac{\omega_r^2}{k^2 v_{Tsd}^2}\right) + \frac{v_{Tsd}^3 \omega_{psi}^2}{v_{Tsi}^3 \omega_{psd}^2} + \frac{v_{Tsd}^3 \omega_{pse}^2}{v_{Tse}^3 \omega_{psd}^2}. \tag{26}$$

The waves are stable if $\gamma < 0$, but the waves are unstable if $\gamma > 0$. And then from Eq.(23) we can find the stability condition,

$$v_{f0} < v_\phi \left(1 + \frac{D'}{C'}\right). \tag{27}$$

There is also a critical velocity of the flowing dusty plasma for the dust-acoustic wave instability, $u'_{f0} \equiv v_\phi (1 + D'/C')$. Namely, if $v_{f0} > u'_{f0}$, the dust-acoustic waves are unstable, in reverse, if $v_{f0} < u'_{f0}$, the waves are stable, but if $v_{f0} = u'_{f0}$, the waves are at the critical stability.

## IV. NUMERICAL ANALYSES OF DUST-ACOUSTIC WAVE INSTABILITY IN THE PERMEATING DUSTY PLASMA

In order to illustrate our theoretical results more clearly, we apply our formulae to an example of the permeating dusty plasma that a cometary plasma tail is passing through the interplanetary space plasma, and then carry out the numerical calculations. In this model of permeating dusty plasma, the cometary plasma tail is the flowing dusty plasma, while the interplanetary medium is the target dusty plasma. The values of physical quantities of the cometary plasma tail can be taken[9,34] as $T_{fe} = 1.16 \times 10^5$ K, $T_{fi} = 2.32 \times 10^4$ K, $T_{fd} = 1.16 \times 10^2$ K, $m_{fi} = 1.67 \times 10^{-27}$ kg $m_{fd} = 1.13 \times 10^{-20}$ kg $n_{fd0} = 10 \text{ m}^{-3}$, $n_{fi0} \approx n_{fe0} = 10^7$ m$^{-3}$ and $Q_{fd}$=816$e$. The physical parameters of the interplanetary medium plasma can be taken[35] as $T_{se} = 2 \times 10^5$ K, $T_{sd} = 200$ K, $n_{se0} = 5 \times 10^6 \text{ m}^{-3}$ and $Q_{sd}(a_0 = 0.1\mu m)$=240$e$. If there is $n_{si0} = n_{se0}$ and $m_{se} \ll m_{si}$, one can take $m_{si} = 2 \times 10^{-26}$ kg. If the dust gains in the interplanetary medium are the same as those in the cometary plasma tail, one can take $m_{sd} = 1.26 \times 10^{-19}$ kg, and

$n_{sd0} = 16\,\text{m}^{-3}$.

Corresponding to the theoretical formulae, now we give our numerical results in the following two limiting physical cases (a) and (b).

(*a*). The case of $v_\phi - v_{f0} \ll v_{Tfd}$, i.e. the thermal velocity of the flowing dusty plasma (i.e. the cometary plasma tail) is much larger than the phase velocity of dust-acoustic waves. After the above example of permeating dust plasma is taken into consideration, three parameters in Eq.(15), given by Eqs.(16)-(18), now are respectively

$$B = \frac{\sqrt{\pi}\omega_r^3}{\omega_r^2 + 6k^2 \frac{k_B T_{sd}}{m_{sd}}} \left( \frac{\omega_r}{k} \sqrt{\frac{m_{sd}}{2k_B T_{sd}}} \right)^3, \tag{28}$$

$$C \approx \frac{n_{fd0} Q_{fd}^2 m_{fd}^{\frac{1}{2}} T_{sd}^{\frac{3}{2}}}{n_{sd0} Q_{sd}^2 m_{sd}^{\frac{1}{2}} T_{fd}^{\frac{3}{2}}} + \frac{n_{fi0} Q_{fi}^2 m_{fi}^{\frac{1}{2}} T_{sd}^{\frac{3}{2}}}{n_{sd0} Q_{sd}^2 m_{sd}^{\frac{1}{2}} T_{fi}^{\frac{3}{2}}}, \tag{29}$$

and

$$D \approx \exp\left(-\frac{\omega_r^2 m_{sd}}{2k^2 k_B T_{sd}}\right) + \frac{n_{si0} Q_{si}^2 m_{si}^{\frac{1}{2}} T_{sd}^{\frac{3}{2}}}{n_{sd0} Q_{sd}^2 m_{sd}^{\frac{1}{2}} T_{si}^{\frac{3}{2}}}, \tag{30}$$

where those terms containing electron's mass have been neglected since they are very small as compared with the terms containing ion's mass and dust grain's mass.

According to Eq.(19) with Eqs.(29) and (30), the instability critical flowing velocity $u_{f0}$ (threshold velocity) of dust-acoustic waves can be illustrated by Fig.1, where $u_{f0}$ is a function of the wave number and appears under the conditions of three different values of the charges, $Q_{sd}$ and $Q_{fd}$, Fig.1 shows that the instability critical flowing velocity $u_{f0}$ is very small ($\sim 10^{-2}$ m s$^{-1}$), and therefore the dust-acoustic waves should be generally unstable. In addition, according to Eq.(15) with the three parameters given by Eqs.(28)-(30) in the case of $v_\phi - v_{f0} \ll v_{Tfd}$, in Fig.2 we show that the instability growth rate $\gamma$ is always more than zero for all the wave number and for any values of the flowing velocity of the cometary plasma tail, and thus the dust-acoustic waves in the permeating dusty plasma are generally unstable.

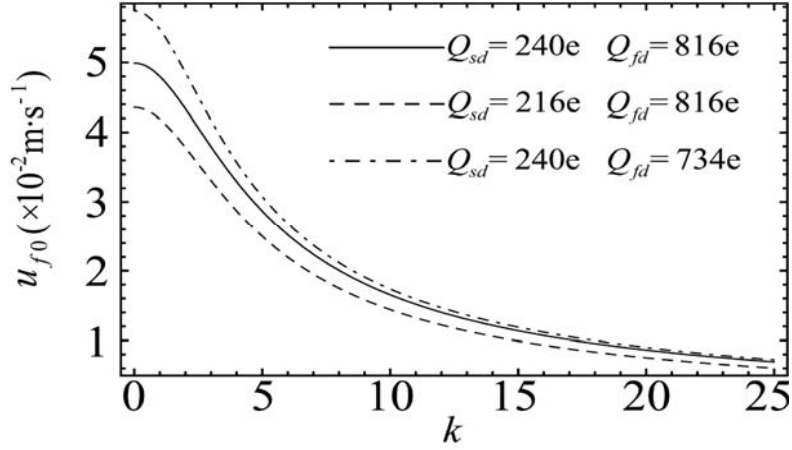

Fig.1. Case (a): the instability critical flowing velocity $u_{f0}$ of dust-acoustic waves for three classes of different values of $Q_{sd}$ and $Q_{fd}$.

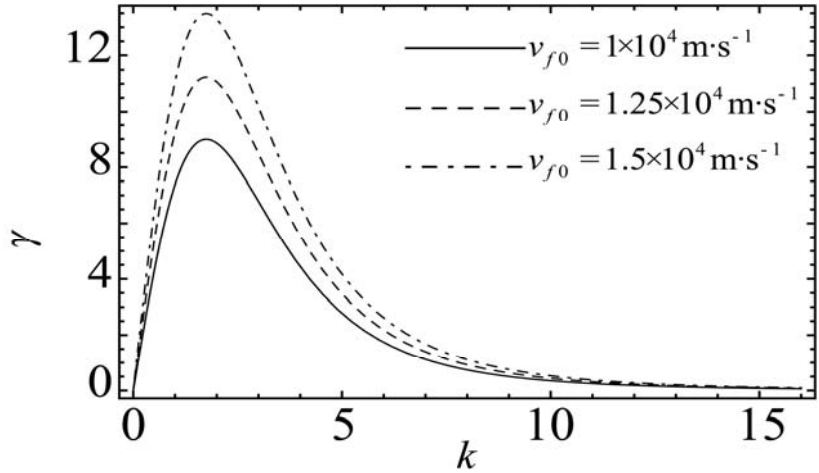

Fig.2. Case (a): the instability growth rate $\gamma$ of dust-acoustic waves for three different values of the flowing velocity: $v_{f0}$.

(b). The case of $v_{Tfd} \ll v_\phi - v_{f0}$, i.e. the thermal velocity of the flowing dusty plasma (i.e. the cometary plasma tail) is much smaller than the phase velocity of the waves. After the above example of permeating dust plasma is taken into consideration, three parameters given by Eqs.(24)-(26) in Eq.(23) now are respectively

$$B' = \frac{\sqrt{\pi}\left(\omega_r^2 \sqrt{m_{sd}/2k^2 k_B T_{sd}}\right)^3}{\omega_r^2 + \frac{6k^2 k_B T_{sd}}{m_{sd}} + \frac{n_{fd} Q_{fd}^2 m_{sd} \omega_r^3}{n_{sd} Q_{sd}^2 m_{fd} (\omega_r - kv_{f0})^3}\left[1 + \frac{6k^2 k_B T_{fd}}{(\omega_r - kv_{f0})^2 m_{fd}}\right]}, \quad (31)$$

$$C' \approx \frac{n_{fi0} Q_{fi}^2 m_{fi}^{\frac{1}{2}} T_{sd}^{\frac{3}{2}}}{n_{sd0} Q_{sd}^2 m_{sd}^{\frac{1}{2}} T_{fi}^{\frac{3}{2}}}, \quad (32)$$

and

$$D' \approx \exp\left(-\frac{\omega_r^2 m_{sd}}{2k^2 k_B T_{sd}}\right) + \frac{n_{si0} Q_{si}^2 m_{si}^{\frac{1}{2}} T_{sd}^{\frac{3}{2}}}{n_{sd0} Q_{sd}^2 m_{sd}^{\frac{1}{2}} T_{si}^{\frac{3}{2}}}. \quad (33)$$

where those terms containing electron's mass have been neglected because they are very small as compared with these terms containing ion's mass and dust grain's mass.

According to Eq.(27) with Eqs.(32) and (33), the instability critical flowing velocity $u'_{f0}$ (threshold velocity) of the dust-acoustic waves is illustrated in Fig.3 as a function of the wave number for two different values of the charge, $Q_{sd}$. Fig.3 shows that the instability critical flowing velocity $u'_{f0}$ is generally large ($\sim 10^4$ m s$^{-1}$), but it is small only when the wave number $k$ is near zero; it increases as wave number $k$ increases, and reaches its peak value when wave number $k$ is about 1, and then decreases gradually as wave number $k$ increases.

According to Eq.(23) with three parameters Eqs.(31)-(33), the instability growth rate $\gamma$ of dust-acoustic waves is illustrated in Fig.4 as a function of the wave number for three different values of the flowing velocity, $v_{f0}$. Fig.4 shows that the dust-acoustic waves in the permeating dusty plasma are almost always stable because the instability growth rate $\gamma$ is always less than zero for almost all the wave number. Only when the flowing velocity of the cometary plasma tail is large and the wave number $k$ is small, might the instability growth rate $\gamma$ be more than zero and thus the dust-acoustic waves is unstable.

When the physical conditions are in between the two limiting physical cases (a) and (b), the numerical results can show a significant change of the dust-acoustic wave stability properties and give a strong insight into the dependence of the dust-acoustic wave instability on such physical conditions as well as the velocity of the flowing dusty plasma (i.e. the cometary plasma tail) in the permeating dusty plasma.

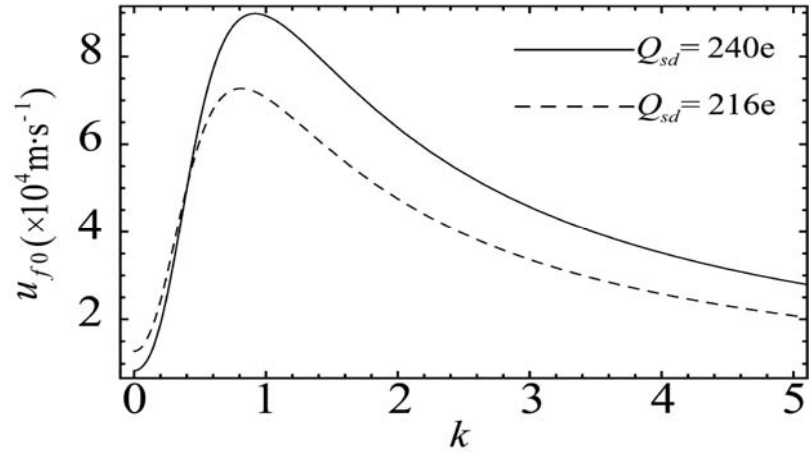

Fig.3. Case (b): the instability critical flowing velocity $u'_{f0}$ of dust-acoustic waves for two different values of the charge $Q_{sd}$.

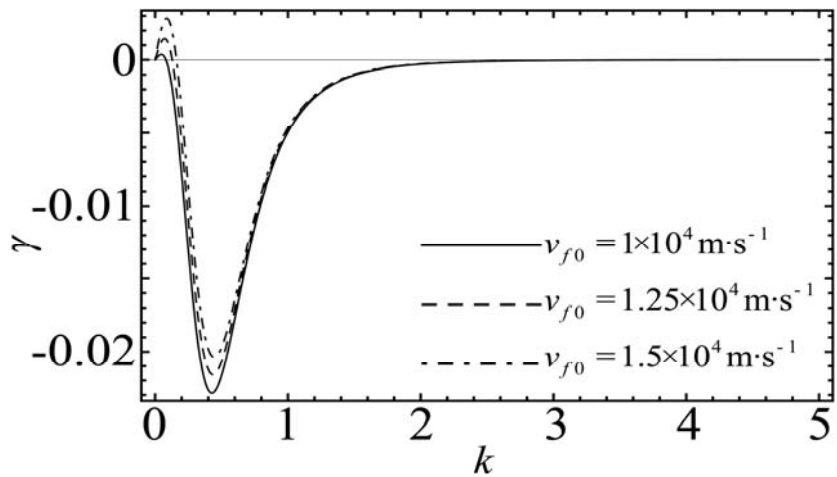

Fig.4. Case (b): the instability growth rate $\gamma$ of dust-acoustic waves for three different values of the flowing velocity: $v_{f0}$.

## V. CONCLUSIONS

In conclusions, the dust-acoustic waves and their stability in the permeating dusty plasma with the Maxwellian velocity distribution have been investigated. We derived the dust-acoustic wave frequency and instability growth rate in two limiting physical cases that the thermal velocity of the flowing dusty plasma is much larger than, and much smaller than the phase velocity of dust-acoustic waves. And therefore we obtained the stability criterions of dust-acoustic waves, which strongly depends on the velocity $v_{f0}$ of the flowing dusty plasma (see Eq.(19) and Eq.(27)) in the permeating dusty plasma. We have analyzed the critical flowing velocity (threshold velocity) for the dust-acoustic wave instability.

In order to illustrate our theoretical results more clearly, we have applied our formulae to an example of permeating dusty plasma that a cometary plasma tail is passing through the interplanetary space, and have made numerical calculations on the dust-acoustic wave instability. The numerical results show that, if the thermal velocity of the flowing dusty plasma (i.e. the cometary plasma tail) is much larger than the phase velocity of dust-acoustic waves, the waves are generally unstable because the instability critical flowing velocity is very small and the instability growth rate $\gamma$ is always more than zero for all the wave number. However, if the thermal velocity of the flowing dusty plasma (i.e. the cometary plasma tail) is much smaller than the phase velocity of dust-acoustic waves and if the flowing velocity is not large, the waves are almost always stable because the instability growth rate $\gamma$ is always less than zero for almost all of the wave number. Only when the wave number is smaller and the flowing velocity of the cometary plasma tail is larger, might the instability growth rate $\gamma$ be more than zero and so the waves be unstable. We have illustrated the critical flowing velocity for the dust-acoustic wave instability as a function of the wave number. When the physical conditions are between these two limiting physical cases, our numerical results gained a strong insight into the dependence of the dust-acoustic wave instability on the physical conditions and the flowing velocity of the cometary plasma tail in the permeating dusty plasma.

ACKNOWLEDGMENTS

This work is supported by the National Natural Science Foundation of China (NSFC) under Grants Nos.11175128 and 10675088.